\documentclass[sigplan,nonacm,screen]{acmart}
\usepackage{amsfonts}
\usepackage{algorithmic}
\usepackage{algorithm}
\usepackage{array}
\usepackage{textcomp}
\usepackage{stfloats}
\usepackage{url}
\usepackage{verbatim}
\usepackage{graphicx}
\usepackage{textcomp}
\usepackage{multicol}
\usepackage{multirow}
\usepackage{siunitx}
\usepackage{comment}
\usepackage{bm}
\usepackage{wrapfig}
\usepackage{enumitem}
\usepackage{cleveref}
\usepackage[normalem]{ulem}
\usepackage{rotating}
\usepackage{hyperref}
\usepackage{booktabs}
\usepackage{balance}
\usepackage{amsmath}
\usepackage{fancyhdr}
\usepackage{fontawesome5}
\usepackage{bbm}
\usepackage{soul}
\usepackage{mathtools}
\usepackage{pifont}
\usepackage{makecell}
\usepackage{subcaption}
\usepackage{xspace}
\usepackage[skins]{tcolorbox}
\usepackage{etoolbox} 

\setlist{noitemsep, leftmargin=*, topsep=0pt, partopsep=0pt}
\crefformat{section}{\S#2#1#3} 
\crefformat{subsection}{\S#2#1#3}
\crefformat{subsubsection}{\S#2#1#3}

\copyrightyear{2026}
\acmYear{2026}
\acmDOI{XXXXXXX.XXXXXXX}
\acmConference[ASPLOS’26]{the ACM International Conference on Architectural Support for Programming Languages and Operating Systems}{March 22-26th, 2026}{Pittsburgh, USA}
\acmISBN{978-1-4503-XXXX-X/2025/06}

\tcbset{
  takeaway/.style={ enhanced,width=\hsize,left=0pt,right=0pt,top=0pt,bottom=0pt,colback=black!20!green!8!white,boxrule=1pt,colframe=black!30!green!30!white
  },
}
\newcounter{takeawaycounter}

\tcbset{
  question/.style={ enhanced,width=\hsize,left=0pt,right=0pt,top=0pt,bottom=0pt,colback=black!20!red!8!white,boxrule=1pt,colframe=black!30!red!30!white
  },
}
\newcounter{questioncounter}

\colorlet{minenergy}{green!20}
\colorlet{lowenergy}{yellow!40}
\colorlet{midenergy}{orange!30}
\colorlet{highenergy}{red!30}
\colorlet{nodata}{gray!50}
\definecolor{mygray}{gray}{0.9}

\newcommand{\cbox}[2][mygray]{%
  \colorbox{#1}{\parbox{\dimexpr\linewidth-2\fboxsep}{\strut #2\strut}}%
}

\setcopyright{none}

\newcommand{\PROFILER}{qMeter}
\newcommand{\FULLPREC}{FP16}
\newcommand{\WEIGHTONLYINT}{W8A16-INT}
\newcommand{\SMOOTHQUANT}{W8A8-INT}
\newcommand{\FPEight}{W8A8-FP}
\newcommand{\FPEightKV}{W8A8KV8-FP}
\newcommand{\AWQWO}{W4A16-INT}
\newcommand{\AWQAC}{W4A8}
\newcommand{\AWQACKV}{W4A8KV8}
\newcommand{\QSERVE}{W4A8KV4}

\begin{document}

\title{Systematic Characterization of LLM Quantization: A Performance, Energy, and Quality Perspective}

\author{Tianyao Shi}
\orcid{0009-0006-6782-0144}
\affiliation{%
  \institution{Purdue University}
   \city{West Lafayette}
   \state{IN}
   \country{USA}}
\email{shi676@purdue.edu}

\author{Yi Ding}
\orcid{0000-0003-2757-9182}
\affiliation{%
   \institution{Purdue University}
   \city{West Lafayette}
   \state{IN}
   \country{USA}}
\email{yiding@purdue.edu}

\begin{abstract}

Large language models (LLMs) have demonstrated remarkable capabilities across diverse domains, but their heavy resource demands make \emph{quantization}—reducing precision to lower-bit formats—critical for efficient serving. While many quantization methods exist, a systematic understanding of their performance, energy, and quality tradeoffs in realistic serving conditions remains a gap.

In this work, we first develop a fully automated online characterization framework \PROFILER{}, and then conduct an in-depth characterization of 11 post-training LLM quantization methods across 4 model sizes (7B–70B) and two GPU architectures (A100, H100). We evaluate quantization at the application, workload, parallelism, and hardware levels under online serving conditions. Our study reveals highly task- and method-dependent tradeoffs, strong sensitivity to workload characteristics, and complex interactions with parallelism and GPU architecture. We further present three optimization case studies illustrating deployment challenges in capacity planning, energy-efficient scheduling, and multi-objective tuning.
To the best of our knowledge, this is one of the first comprehensive application-, system-, and hardware-level characterization of LLM quantization from a joint performance, energy, and quality perspective.

\end{abstract}

\maketitle

\section{Introduction}


Large language models (LLMs)~\cite{chatgpt,deepseek,Gemini,meta_llama} have demonstrated remarkable capabilities across a wide range of tasks such as software development~\cite{he2025llm,copilot}, scientific discovery~\cite{zheng2025large}, and healthcare~\cite{qiu2024llm,Peng2023}. Due to their large size, \emph{quantization}—reducing model precision from full precision to lower-bit formats (e.g., 8-bit, 4-bit)—has been widely adopted to shrink model size and improve serving efficiency~\cite{gong2025survey}. In recent years, numerous quantization techniques have emerged, including \emph{weight-only} (only weights are quantized into
lower bits)~\cite{frantar2022gptq,dettmers2023case,sheng2023flexgen,park2022lut,lin2024awq} and \emph{activation} (both activation and weights are
quantized)~\cite{dettmers2022gpt3,xiao2023smoothquant,yao2022zeroquant,wei2022outlier,wang2023peeling,ashkboos2024quarot} quantization, with more recent approaches incorporating KV cache compression for additional size reduction~\cite{lin2024qserve}.

Despite the broad adoption of quantization and extensive characterization on LLM serving~\cite{patel2024characterizing,hu2024characterization,stojkovic2025dynamollm,nguyen2024towards,zhong2024distserve,patel2024splitwise,lazuka2024llm}, systematic characterization of LLM quantization remains underexplored. While recent studies have begun to address this gap~\cite{kurtic2024give,lee2024exploring,de2024exploration}, they exhibit the following limitations:

\begin{itemize}
    \item \textbf{Lack of performance-energy-quality triad analysis.} Existing LLM quantization studies evaluate only partial tradeoffs: some focus solely on a single metric such as output quality~\cite{lee2024exploring} or performance (latency/throughput), while others examine two metrics in combination, e.g., accuracy–performance~\cite{kurtic2024give} or energy–performance~\cite{de2024exploration}. None evaluate performance, energy, and quality together, leaving the full design tradeoff space unexplored and limiting informed deployment decisions.
    \item \textbf{Lack of online characterization.} Existing LLM quantization characterization work focuses on offline profiling using fixed prompts under controlled conditions rather than online setting, which fails to capture how quantized models behave in dynamic, real-world serving scenarios with varying load and input distributions. 
    \item \textbf{Lack of system-level optimization.} Existing LLM quantization characterization studies are mostly from the machine learning community, overlooking advanced system-level optimization techniques such as parallelism (for scaling model serving across multiple devices) and KV cache compression (for further model size reduction). This makes it unclear how quantization interacts with distributed execution and memory management.
    \item \textbf{Limited system design insights.} Existing LLM quantization studies largely focus on application-level metrics such as latency, throughput, or quality. They rarely translate these findings into implications for architectural and system design, deployment strategies, and resource allocation in production environments, where performance, energy, quality, and service-level objectives (SLOs) must be explicitly managed.
\end{itemize}

To overcome these limitations, we conduct a comprehensive, online systematic characterization of LLM quantization, evaluating performance, energy efficiency, and output quality jointly across diverse model sizes, quantization methods, and workloads. To facilitate accurate and repeatable online profiling, we develop \PROFILER{} (\Cref{sec:method}), a fully automated online characterization framework that detects saturation points, sweeps large configuration spaces (quantization schemes, parallelism levels, workloads), and integrates with inference engines and benchmarking suites. \PROFILER{} ensures measurement robustness by continuously monitoring serving engine health and restarting failed instances, enabling consistent profiling across diverse load conditions.

Using \PROFILER{}, we conduct an in-depth characterization of 11 LLM post-training quantization methods (summarized in \Cref{tab:quant_method_scope}) on the TensorRT-LLM v0.19.0 inference engine~\cite{TensorRT-LLM}. Our experiments cover the Llama-2 model family~\cite{touvron2023Llama2} at four sizes (7B, 13B, 34B, and 70B), evaluated on NVIDIA H100 and A100 GPUs. Following prior work~\cite{zhong2024distserve,stojkovic2025dynamollm}, we include chatbot, code generation, and summarization tasks across diverse benchmarks. The study examines quantization at four levels: application, workload, parallelism, and hardware. Our results reveal the following key insights:

\begin{itemize}
    \item \textbf{Task- and method-dependent tradeoffs (\Cref{sec:char-app}).}  No single quantization method dominates across latency, energy efficiency, and quality. Larger quantized models can sometimes outperform smaller full-precision ones, but improvements are highly task- and method-dependent.
    \item  \textbf{Workload Sensitivity (\Cref{sec:char-workload}).} Quantization benefits vary with input/output length and load intensity: short outputs can hurt TTFT, long inputs can increase TPOT, and optimal configurations shift with request rate.
    \item \textbf{Parallelism interaction (\Cref{sec:char-par}).} Activation quantization scales with moderate tensor parallelism (TP), even reducing GPU needs. Weight-only with KV compression incurs compounded latency and energy overhead. This indicates that quantization and parallelism should be co-optimized.
    \item \textbf{Hardware dependence (\Cref{sec:char-hw}).} GPU architecture affects quantization tradeoffs: H100 improves latency and scalability, while A100 offers higher energy efficiency at moderate loads; memory and compute capacity jointly shape saturation behavior.
\end{itemize}

Building on the characterization results, we present three optimization case studies for quantization-enabled LLM serving clusters, motivated by real-world deployment needs and guided by model–system–hardware co-design (\Cref{sec:opt}). The first examines saturation point prediction for scheduling and capacity planning. The second explores energy-optimal configuration by examining how different parallelization strategies impact efficiency under varying traffic loads. The third highlights the energy–quality imbalance in single-objective optimization, where prioritizing one metric can degrade the other. Together, these case studies demonstrate the need for holistic approaches that balance performance, energy, and quality across models, systems, and hardware.

To the best of our knowledge, this is one of the first comprehensive application-, system-, and hardware-level characterization of LLM quantization from a joint performance, energy, and quality perspective. We hope to lay the foundation for future research on principled model, system, hardware co-design for quantization-enabled LLM serving at scale.

\section{Background and Related Work}\label{sec:background} 


In this section, we first review the LLM quantization techniques analyzed in this paper, followed by background on recent LLM characterization studies to highlight the gap in quantization-focused characterization.

\subsection{LLM Quantization}

\begin{table}[]
    \footnotesize
    \centering
    \caption{Quantization methods studied in this paper.}
    \label{tab:quant_method_scope}
    \resizebox{\linewidth}{!}{
    \begin{tabular}{lll}
    \toprule
        \textbf{Category} &  \textbf{Method} & \textbf{Name}\\
        \midrule
         Weight &  Per-Channel INT8~\cite{jacob2018quantization} & \WEIGHTONLYINT{} \\
         Only & AWQ~\cite{lin2024awq} & \AWQWO{} \\
         \midrule
         & SmoothQuant~\cite{xiao2023smoothquant} & \SMOOTHQUANT{} \\
        Activation & Per-Tensor FP8\cite{micikevicius2022fp8} & \FPEight{} \\
         
        & AWQ & \AWQAC{} \\ 
        \midrule
         & QServe~\cite{lin2024qserve} & \QSERVE{} \\    
        KV Cache & - & W8A16KV8-INT, W4A16KV8-INT,\\
        Compression & - & W8A8KV8-INT, W8A8KV8-FP \\ 
        & - & W4A8KV8 \\
        \bottomrule
    \end{tabular}}
    
\end{table}

Quantization techniques can be broadly categorized into \emph{quantization-aware training} (QAT), which incorporates quantization into the training process via backpropagation to update quantized weights~\cite{choi2018pact,nagel2021white}, and \emph{post-training quantization} (PTQ), which is typically training-free~\cite{nagel2019data,nagel2020up}. Since QAT is difficult to scale to large models such as LLMs, PTQ is the dominant approach for LLM quantization and is the also focus of this study.

There are three major PTQ techniques. The first is \emph{weight-only} quantization, where only weights are quantized into low-bit integers~\cite{frantar2022gptq,dettmers2023case,sheng2023flexgen,park2022lut,lin2024awq}. 
For this category, we study per-channel INT8~\cite{jacob2018quantization} and AWQ~\cite{lin2024awq} in this paper. 
The second is \emph{activation} quantization, where both activation and weights are quantized to INT8~\cite{dettmers2022gpt3,xiao2023smoothquant,yao2022zeroquant,wei2022outlier,wang2023peeling,ashkboos2024quarot}. Activation quantization generally outperforms weight-only quantization as it reduces memory requirements while also accelerating token generation in memory-bound workloads. 
For this category, we study SmoothQuant~\cite{xiao2023smoothquant}, per-tensor FP8~\cite{micikevicius2022fp8}, and W4A8-AWQ methods in this paper. 
The third is \emph{KV cache compression}, which quantizes KV cache along with activation and weights~\cite{lin2024qserve}. For this category, we study QServe and 8-bit KV cache compression variant of all the \emph{weight-only} and \emph{activation} method mentioned above. 
All quantization methods considered in this work are widely adopted in real-world production and are supported by high-performance inference engines such as TensorRT-LLM.



\subsection{LLM Characterization}

Recent profiling studies have characterized LLM serving to understand the complex interplay between model architectures, inference workloads, and system-level characteristics. Such work provides empirical insights into performance, scalability, latency, and energy efficiency to inform optimization opportunities. On the machine learning side, many studies benchmark LLMs across various domain-specific tasks to measure accuracy and speed~\cite{mishra2024characterizing,poddar2025brevity,cheng2023compost,wen2024characterizing,xu2024do,ko2024on,wei2022emergent}. On the systems side, characterization efforts focus on latency~\cite{hu2024characterization,zhong2024distserve,lazuka2024llm}, especially detailed latency breakdowns for TTPT during the prefill phase and TPOT during decoding. Beyond latency, energy efficiency and the associated trade-offs are also commonly analyzed~~\cite{patel2024characterizing,stojkovic2025dynamollm,patel2024splitwise}. More recently, there has been growing interest in assessing environmental impacts of LLM serving such as carbon emissions~\cite{nguyen2024towards,wu2025fuel} and water consumption~\cite{wu2025scarf}. 

Despite extensive characterization of general LLM serving, only a limited number of studies have examined LLM quantization. Existing quantization characterization has evaluated only partial tradeoffs: some focus exclusively on a single metric such as accuracy~\cite{lee2024exploring} or performance (e.g., latency or throughput), while others investigate pairs of metrics like accuracy and performance~\cite{kurtic2024give} or energy and performance~\cite{de2024exploration}. However, none have comprehensively considered all three metrics together. However, none have jointly examined all three metrics. In addition, existing studies rely on offline profiling, which cannot capture quantization behavior under dynamic, real-world online serving conditions. These gaps limit our understanding of the full design tradeoffs in LLM quantization and also motivate this work.
\section{Methodology}\label{sec:method}

In this section, we first introduce our newly developed tool, \PROFILER{}, which we use for the characterization study in this work, and then describe the testbed configurations for all experiments presented in the paper.

\begin{figure}
    \centering
    \includegraphics[width=0.99\linewidth]{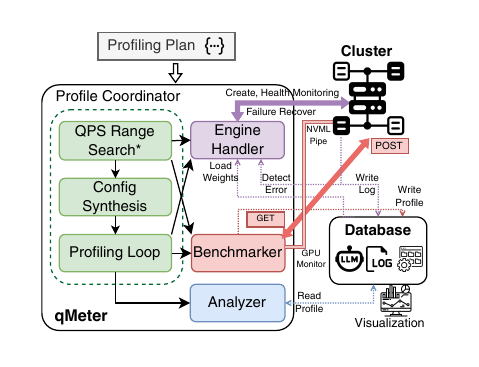}
    \caption{\textbf{The flowchart of \PROFILER{} (\Cref{sec:profiler})}. We build \PROFILER{} that interacts with the GPU cluster and database to run controlled tests and profile performance, energy, and quality metrics. }
    \label{fig:profiler}
\end{figure}

\subsection{\PROFILER}\label{sec:profiler}

In this work, we introduce \PROFILER{}, a tool for automated and comprehensive LLM profiling. While \PROFILER{} is designed for LLM quantization profiling, it generalizes to profiling any LLM workload. We build \PROFILER{} in response to real needs and challenges encountered in our studies:
\begin{itemize}
    \item \textbf{Large number of experiment settings and diverse profiling configurations.} Our study spans a wide range of model sizes, quantization schemes, parallelism strategies, and LLM serving applications. Collecting performance, energy, and quality metrics across all these combinations efficiently and consistently is a complex and time-consuming process. An automated approach is critical—not only to systematically generate and execute these configurations, but also to identify the \emph{saturation point} for each setup, i.e., the maximum load the LLM serving system can sustain before performance plateaus or degrades.
    \item \textbf{Mismatch between existing benchmarking tools and our evaluation goals.} Many existing LLM quantization profiling tools focus on offline or throughput-centric benchmarking~\cite{kurtic2024give,lee2024exploring,de2024exploration}, whereas our evaluation targets online serving scenarios that measure latency, energy consumption, and output quality simultaneously. This mismatch means that existing tools often cannot capture the fine-grained measurements needed for our analysis.
    \item \textbf{Fragile serving engines under extreme workloads.} Under high load or stress-testing conditions, existing serving engines can become unstable, crash, or produce inconsistent results. This instability makes it challenging to run large-scale, high-intensity profiling without robust monitoring and recovery mechanisms.
\end{itemize}

\Cref{fig:profiler} shows \PROFILER{} and how it interacts with the GPU cluster and the database.
The user specifies the mix of model, quantization method, parallelism strategy, dataset or quality benchmark in the profiling plan via a JSON file or command line arguments. Then the Profile Coordinator guides the profiling process by first calling the Engine Handler to create the corresponding LLM serving instance.
For latency and energy profile, it conducts a request rate (QPS, req/s) range search to find the saturation point of each system configuration \emph{(model size, quantization method, GPU type, parallelism, dataset)}. Specifically, it calls the Benchmarker to run short-period bursting benchmarks that interact with the LLM serving instances via HTTP get/post and check if the predefined SLO is violated. The highest SLO-attaining QPS is determined by binary search and deemed as the saturation point. Based on the saturation point, the Profile Coordinator generates the profiling configurations spanning the QPS range and invokes the Benchmarker in the main profiling loop to collect latency and energy measurements. 
For quality profile, \PROFILER{} bypasses the QPS range search stage and directly generates configuration files for benchmarking suites like \verb|lmeval| and \verb|opencompass|, which act as the Benchmarker.

To ensure reliability and avoid redundant runs, the Engine Handler continuously monitors the health of serving instances. Upon detecting that the engine is not responding, it will send signals to pause profiling, inspect logs for errors, and verify if the engine is corrupted. Once confirmed, it will kill the old serving processes and restart the new ones to resume profiling once the new instance is operational.

\subsection{Testbed}\label{sec:char-setup}

\begin{table}[t]
\footnotesize
  \centering
  \caption{Datasets for different tasks.}
  \label{tab:ctx-avg}
  \begin{tabular}{llll}
    \toprule
    \textbf{Dataset / Domain} & \textbf{\#Input}  & \textbf{\#Output}  & \textbf{Mid-Range QPS}\\ \midrule
    ShareGPT (Chat)~\cite{sharegpt}          & 331  & 231 & 5 req/s\\
    HumanEval (Code)~\cite{chen2021evaluating}         & 193  & 67 & 21 req/s\\
    NewsQA (Sum.)~\cite{trischler2016newsqa}     & 806 & 200 & 4 req/s\\ \bottomrule
  \end{tabular}
\end{table}

\noindent \textbf{Configurations.} We evaluate both full precision (\FULLPREC) and 11 quantization methods (summarized in~\Cref{tab:quant_method_scope}) spanning weight-only, activation, and KV cache quantization schemes. To broaden the characterization beyond prior LLM quantization studies~\cite{kurtic2024give,lee2024exploring,de2024exploration}, we also vary parallelism strategies. We focus on tensor parallelism (TP), which partitions LLM layers across multiple GPUs to execute them in parallel. We prioritize TP due to its superior throughput and latency compared to pipeline parallelism in single-node deployments~\cite{stojkovic2025dynamollm}. Since most open-source LLMs can fit within the memory budget of 8 GPUs on a single server, we restrict experiments to TP configurations over 1, 2, 4, and 8 GPUs, denoted TP1, TP2, TP4, and TP8, respectively.

\noindent \textbf{Models.} We experiment with open-source Llama-2~\cite{touvron2023Llama2} models in four sizes: 7B, 13B, 70B, and CodeLlama-34B.

\noindent \textbf{Hardware.} We primary experiment with NVIDIA H100 GPU architecture. For hardware level analysis, we experiment with both H100 and A100 to capture performance and energy trade-offs across hardware generations.

\noindent \textbf{Inference engine.} We use TensorRT-LLM v0.19.0~\cite{TensorRT-LLM} due to its native support of diverse quantization methods, integration with CUDA kernels optimized for low-latency inference, and broad adoption in both research and production.

\noindent \textbf{Workloads.} Following prior work~\cite{zhong2024distserve,stojkovic2025dynamollm}, our workload suite includes chatbot, code generation, and summarization tasks, which evaluate on ShareGPT~\cite{sharegpt}, HumanEval~\cite{chen2021evaluating}, and NewsQA~\cite{trischler2016newsqa} datasets respectively. The details of datasets are summarized in~\Cref{tab:ctx-avg}. These datasets are used for evaluating performance, energy efficiency, and output quality. The only exception is that for evaluating output quality in the chatbot task, ShareGPT is not appropriate since it consists of open-ended multi-turn conversations without standardized groundtruth responses, making objective quality metrics infeasible. Instead, we use the following benchmarks to assess chatbot quality, organized into three categories: (1) \texttt{chat-S} for commonsense knowledge and general QA (HellaSwag~\cite{zellers2019hellaswag}, ARC-C~\cite{clark2018think}, Winogrande~\cite{sakaguchi2021winogrande}, TriviaQA~\cite{joshi2017triviaqa}); (2) \texttt{chat-R} for complex instruction following and reasoning (BigBench-Hard~\cite{suzgun2022challenging}, MMLU~\cite{hendryckstest2021}); and (3) \texttt{chat-M} for math problems (GSM8K~\cite{cobbe2021training}, GPQA Diamond~\cite{rein2024gpqa}). 

\noindent \textbf{Metrics.} We quantify performance using Time to First Token (TTFT) and Time per Output Token (TPOT), both measured at various percentile latencies (e.g., P50, P90). Energy efficiency is evaluated via energy per token (Joule/token) using GPU power telemetry. Output quality is assessed via accuracy for chatbot tasks, pass$@$1 for code generation~\cite{chen2021evaluating}, and ROUGE scores for summarization~\cite{lin2004rouge}.

\section{Application Level Analysis}\label{sec:char-app}

In this section, we characterize application-level behaviors by first examining the impact of different quantization methods on latency, energy efficiency, and output quality independently under a fixed model–hardware configuration. We then analyze tradeoffs among these three metrics.

We quantize Llama-2 (7B, 13B, 70B) and CodeLlama-34B using TensorRT-LLM’s quantization utilities. Models up to 34B are deployed on a single H100 GPU, while the 70B model requires TP4, the minimum number of H100s for \FULLPREC{} inference. We replay a mid-range request stream defined as 50\% of the saturation throughput of the 13B INT8 model on an H100. This workload represents a consistent, moderate load: high enough to enable continuous batching, but below saturation to keep quantization effects observable. Task characteristics, including average input/output lengths and mid-range request rates, are summarized in~\Cref{tab:ctx-avg}. We measure TTFT and TPOT over a 2-minute trace, energy per token from H100 telemetry, and output quality (0–100). Due to page limits, we report P90 tail latency results for the 34B model as a representative case, since it is the largest model that fits on a single H100 in \FULLPREC{} and exhibits trends consistent with other model sizes and average latencies.

\subsection{Latency, Energy, and Quality Results}\label{sec:char-prem}

\begin{figure}[t]
  \centering
  \includegraphics[width=0.99\linewidth]{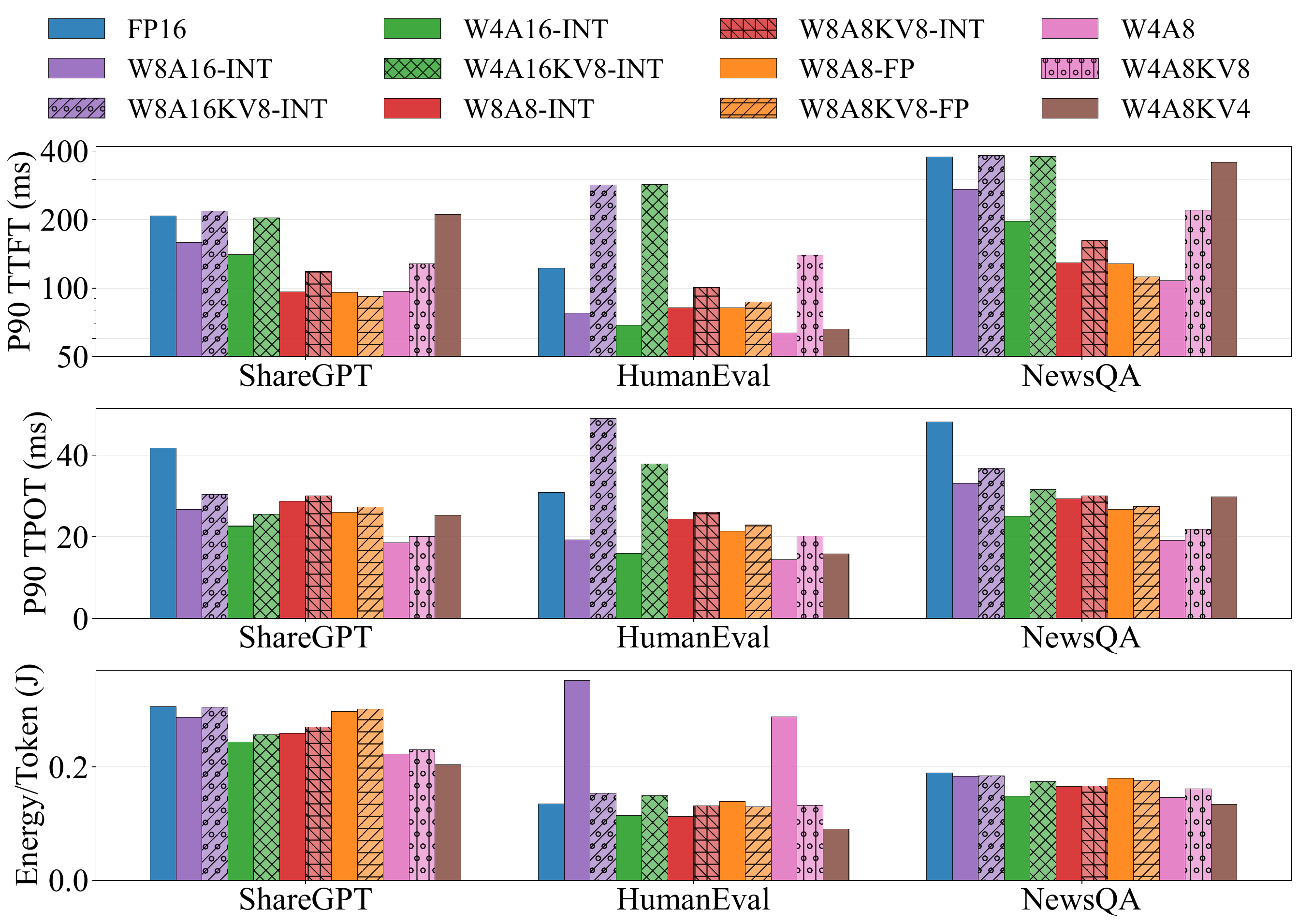}
  \caption{\textbf{Latency and energy efficiency comparison for quantized CodeLlama-34B models (\Cref{sec:char-prem})}.}
  \label{fig:radar-single}
\end{figure}

\begin{figure*}
    \centering
    \includegraphics[width=0.99\textwidth]{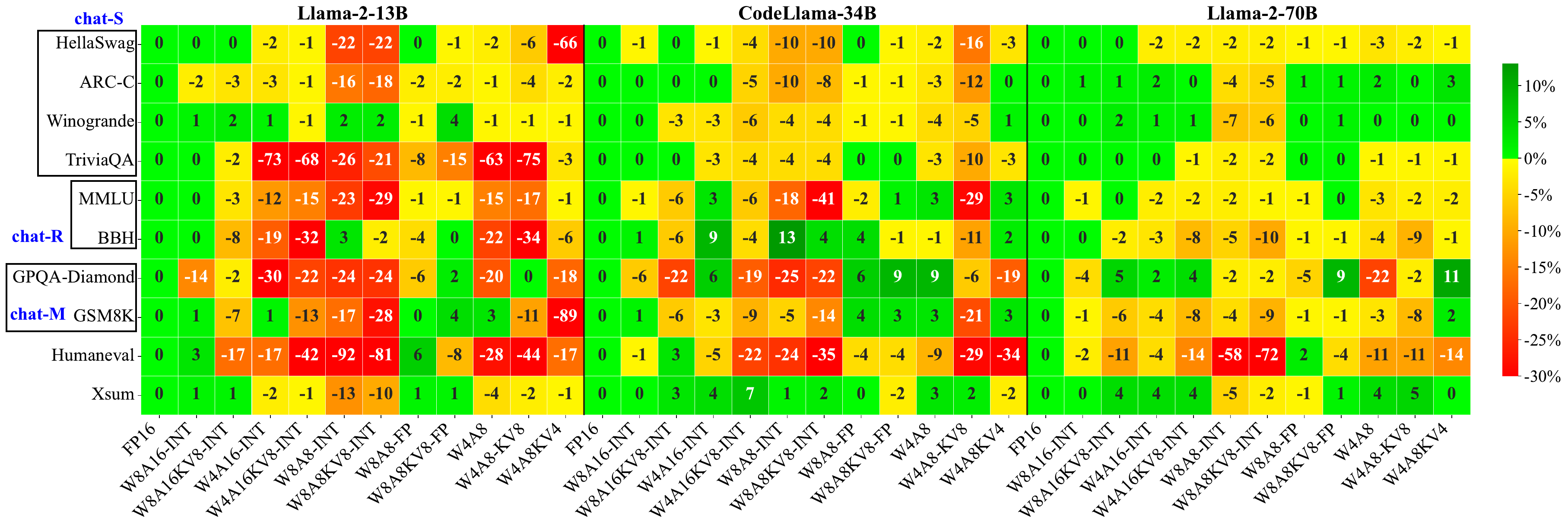}
    \caption{\textbf{Quality score percentage change w.r.t. FP16} across quantization methods for 13B, 34B, and 70B models (\Cref{sec:char-prem}).}
    \label{fig:quality_heatmap}
\end{figure*}

\paragraph{(Tail) Latency.} \Cref{fig:radar-single} (top two rows) reports P90 TTFT and TPOT for each task across quantized 34B models. We make the following observations. \textbf{(1)} \emph{Weight-only} and \emph{activation} quantization methods generally reduce tail latency, with up to a 70\% decrease. Among them, \AWQAC{} consistently delivers the best TTFT and TPOT across all datasets. \textbf{(2)} In general, greater model compression yields better latency: e.g., \AWQWO{} outperforms \WEIGHTONLYINT{}, and \emph{activation} methods (\SMOOTHQUANT{}, \FPEight{}, \AWQAC{}) achieve lower TTFT than \emph{weight-only} methods on ShareGPT and NewsQA, where input sequences are sufficiently long. In contrast, on HumanEval (short contexts), \WEIGHTONLYINT{} and \AWQWO{} surpass \emph{activation} methods, suggesting that context length influences latency behavior (see more in~\Cref{sec:char-context}). \textbf{(3)} Although prior work~\cite{kurtic2024give,lee2024exploring,de2024exploration} reports substantial throughput gains in offline batch mode, our results show that at moderate load, quantization does not always reduce latency in online serving. \textbf{(4)} \emph{KV cache compression} is surprisingly harmful, reducing latency benefits across all methods. Even for \AWQAC{}, applying 8-bit or 4-bit KV cache compression significantly increases latency, and \QSERVE{} shows negligible TTFT gains on ShareGPT and NewsQA. For \emph{weight-only} methods, KV compression can push latency close to \FULLPREC{} levels.

\paragraph{Energy.} \Cref{fig:radar-single} (bottom row) shows energy per token across methods. We make the following observations. \textbf{(1)} Quantization improves energy efficiency by up to 30\%, but 8-bit activation methods yield minimal energy savings despite lower latency. \textbf{(2)} 4-bit quantization delivers larger reductions, with \QSERVE{} achieving the best energy efficiency via combining 4-bit weight and 8-bit activation compression. \textbf{(3)} Exceptions arise for HumanEval again, where \WEIGHTONLYINT{} and \AWQAC{} consume more energy than \FULLPREC{} despite latency gains. \textbf{(4)} KV cache compression generally increases energy consumption, except for \QSERVE{}.

\paragraph{Quality.} \Cref{fig:quality_heatmap} shows normalized benchmark accuracy score changes relative to \FULLPREC{} for 13B, 34B, and 70B models (negative indicates quality loss). Key observations:
\begin{itemize}
    \item \textbf{Efficiency–quality tradeoffs.} Quantization can cause substantial quality degradation, up to a 92\% drop in HumanEval pass rate for the 13B model. Quality-preserving methods include \WEIGHTONLYINT{}, \FPEight{}, and \FPEightKV{}, though these still lose 5–10\% accuracy on some tasks. \AWQWO{} and \AWQAC{} incur up to 22\% loss on larger models, while \SMOOTHQUANT{} suffers the most severe degradation. KV cache compression typically worsens quality for \emph{weight-only} methods and also for \AWQAC{} on 34B.
    \item \textbf{Task difficulty sensitivity.} Quality losses are smaller on simpler QA and summarization tasks, but severe for challenging reasoning tasks like coding and math.
    \item \textbf{Model size sensitivity.} Smaller models are more vulnerable towards quantization-induced quality degradation. This can be confirmed by visually checking the area of red and yellow cells for each model size. For 13B, quality loss is widespread and severe; for 34B, degradation is moderate for \texttt{chat-S} and \texttt{chat-R} and severe in \texttt{chat-M} and coding; for 70B, \texttt{chat-S} remains nearly lossless, with at most 22\% loss for coding and reasoning tasks except \SMOOTHQUANT{}.
\end{itemize}

\begin{figure*}[t]
  \centering
  \includegraphics[width=\textwidth]{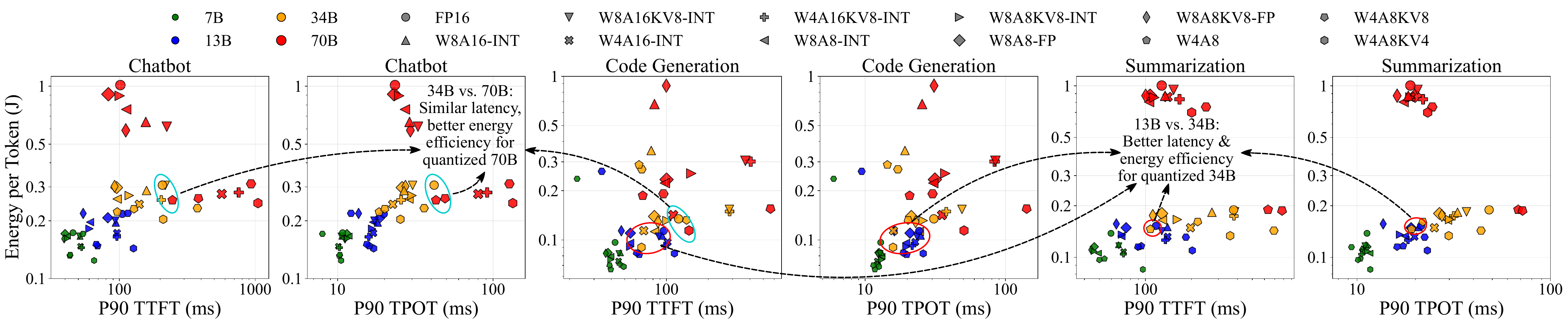}
  \caption{\textbf{Latency vs. energy tradeoffs} across model sizes and quantization methods at mid-range load  (\Cref{sec:char-tradeoff}).}  
  \label{fig:l-e-scatter}
\end{figure*}

\begin{figure*}[t]
  \centering
  \includegraphics[width=\textwidth]{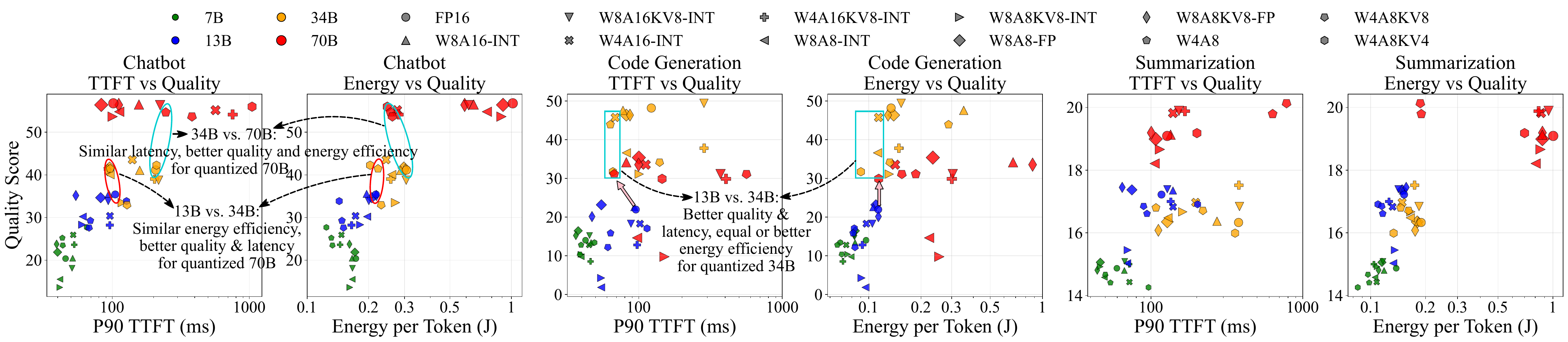}
  \caption{\textbf{Quality vs. latency and energy tradeoffs} across model sizes and quantization methods at mid-range load (\Cref{sec:char-tradeoff}).}
  \label{fig:le-q-scatter}
\end{figure*}

\subsection{Tradeoff Analysis}\label{sec:char-tradeoff}

\paragraph{Latency vs. Energy.} We examine the latency–energy tradeoffs across 4 model sizes and 12 methods on each dataset in~\Cref{fig:l-e-scatter}. The x-axes show P90 TTFT and TPOT, the y-axis shows energy per token at mid-range load, with color and marker size denoting model size and marker type denoting quantization method. Configurations closer to the bottom-left indicate better latency–energy tradeoffs. We want to find if any quantized larger model lies towards left below a smaller \FULLPREC{} model.

Comparing 34B \FULLPREC{} (yellow circle) with nearby quantized 70B models, we find two cases where 70B \AWQAC{} (red point-up pentagon) and \QSERVE{} (red hexagon) have similar latency to 34B \FULLPREC{} with reduced energy per token on chatbot and code generation tasks. Similarly, between 13B \FULLPREC{} (blue circle) and quantized 34B models, we find three cases where 34B \QSERVE{} (yellow hexagon) and \SMOOTHQUANT{} (yellow left triangle) on code completion as well as \AWQAC{} (yellow point-up pentagon) on summarization achieve both lower latency and energy compared to 13B \FULLPREC{}.

\paragraph{Latency and Energy vs. Quality.} We further examine the latency-quality and energy-quality tradeoffs in~\Cref{fig:le-q-scatter}. 
The x-axes show P90 TTFT and energy per token at mid-range load, and the y-axis shows quality scores (chatbot quality is the arithmetic mean of \texttt{chat-R} benchmarks). Higher and more leftward points indicate better tradeoffs. Due to page limits, we present only TTFT results, having checked that TPOT trends are consistent. We want to find if any quantized larger model lies above and left of a smaller \FULLPREC{} model.

Here are the example cases we find. \textbf{(1)} For chatbot workloads, we identify two cross-size tradeoff cases: 34B \AWQAC{} (yellow pentagon) achieves higher output quality and lower latency with only negligible energy overhead compared to 13B \FULLPREC{} (blue circle); and 70B \AWQAC{} (red pentagon) improves both energy efficiency and quality, with only a marginal latency increase. \textbf{(2)} For code generation, 34B \AWQWO{}, \SMOOTHQUANT{}, and \QSERVE{} all outperform 13B \FULLPREC{} by providing higher quality and faster latency without additional energy penalties. In fact, CodeLlama-34B shows superior coding performance even over 70B models, though its summarization accuracy falls slightly below 13B. \textbf{(3)} For summarization, however, no quantized larger model is able to improve quality without incurring penalties in latency or energy efficiency. Overall, these results highlight that the benefits of quantization are highly task- and model-dependent, with non-trivial cross-size tradeoffs that can be exploited in scheduling and capacity planning.


\noindent \cbox{\noindent \textbf{Finding \#1:} \textbf{(a)} No single quantization method dominates across all three metrics, which are varied by task, model size, and precision level. \textbf{(b)} Quantized larger models can match or surpass smaller \FULLPREC{} models in certain tradeoff spaces, offering either better energy with comparable latency or better quality/latency without major energy penalties. \textbf{(c)} Task specialization matters; e.g., CodeLlama-34B significantly outperforms even 70B models on coding tasks but may underperform on summarization. \textbf{(d)} Tradeoff improvements are task- and method-dependent, as gains in quality often come at the expense of latency or energy, with few configurations improving all metrics at once.

\noindent \textbf{Recommendation \#1:} \textbf{(a)} Quantization methods should aim to improve all three metrics together rather than optimizing for one or two metrics, as current approaches rarely achieve simultaneous improvements. \textbf{(b)} Model size and precision should be co-optimized rather than independently. \textbf{(c)} There is a need for adaptive scheduling and model selection strategies that tailor quantization choices to task-specific latency, energy, and quality requirements. \textbf{(d)} Automated tool is needed to navigate the tradeoff spaces and select configurations that best meet application SLOs.
}

\section{Workload Level Analysis}\label{sec:char-workload}

Since our study targets real-world online serving, this section focuses on online workload characteristics and analysis. We will first examine the impact of input/output lengths, and then the effect of load intensity.

\subsection{Input/Output Lengths}\label{sec:char-context}

\begin{figure}[t]
  \centering
  \includegraphics[width=\linewidth]{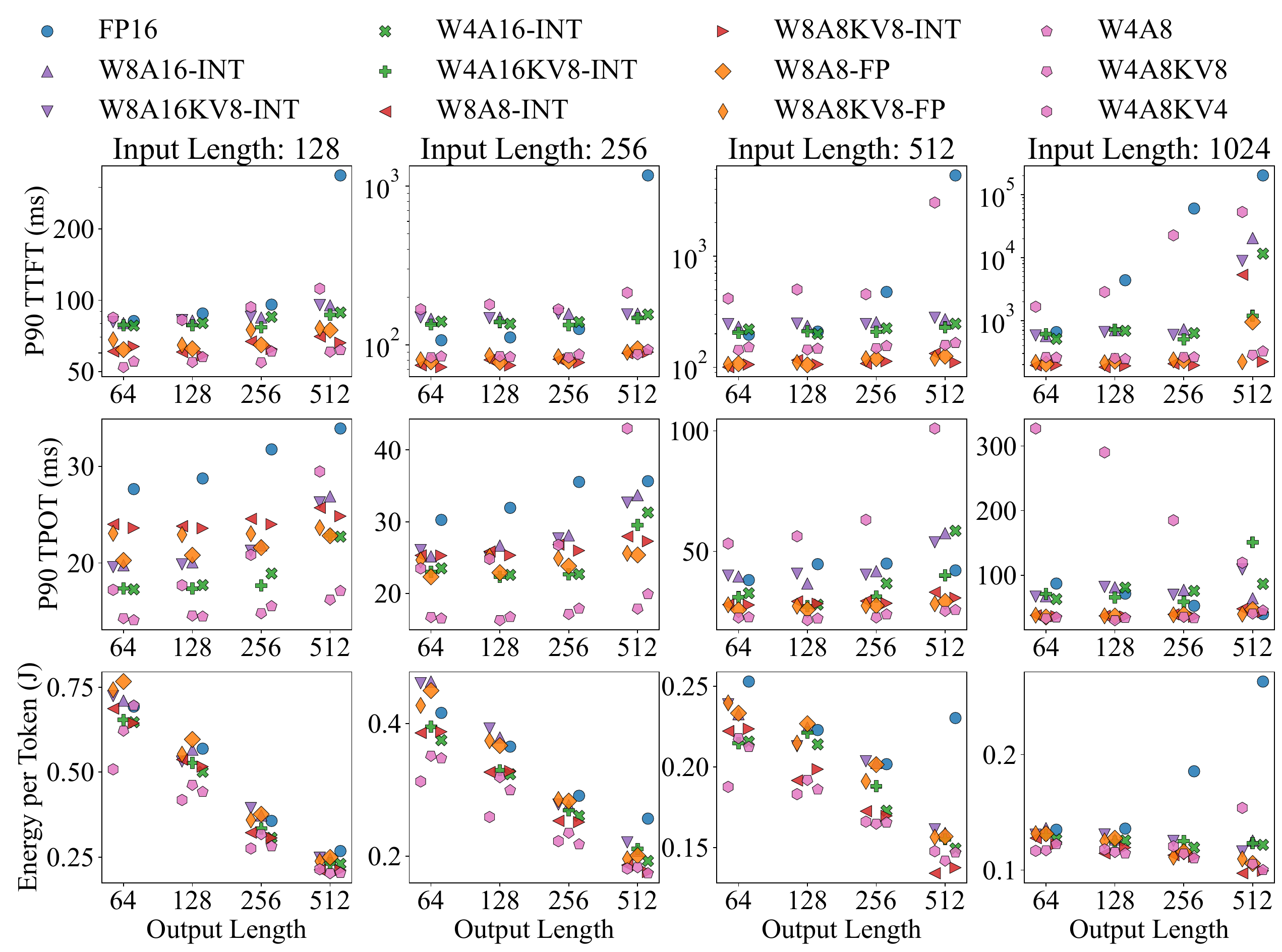}
  \caption{\textbf{Input/output length influence (\Cref{sec:char-context}).} Latency and energy metrics w.r.t. different input/output lengths across quantized 34B models at QPS=5 req/s.}
  \label{fig:context-length}
\end{figure}

\Cref{fig:qps-curves} reports P90 TTFT, P90 TPOT, and energy per token for quantized 34B models across varying input/output lengths. All experiments are run at a fixed QPS of 5 req/s, chosen based on the saturation point of the longest input/output lengths on a single H100. The x-axes represent QPS (req/s), and the y-axes measure latency (ms) or energy per token (J). For visual clarity, the QPS values of different quantization methods are slightly offset, though they correspond to the same request rate. To control input/output lengths, we follow the same methodology from prior work~\cite{stojkovic2025dynamollm}. We set short/medium/long inputs to 128/(256, 512)/1024 tokens, and short/medium/long outputs to 64/(128, 256)/512 tokens. This yields 16 input-output combinations in total.

We make the following observations. \textbf{(1)} TTFT can degrade for some \emph{weight-only} methods when output length is short. This can be observed in input–output pairs such as (128,64), (256,64/128), (512,64/128), and (1024,64). This indicates that the overhead of weight dequantization and prefill computation dominates the total runtime and is not well amortized when the decoding phase is short, making TTFT less efficient. \textbf{(2)} TPOT can degrade when input length is long. This can be observed in \QSERVE{}, \WEIGHTONLYINT{}, \AWQWO{}, and W8A16KV8-INT with 512/1024-token inputs. This reflects increased inefficiency from quantization in handling larger activation matrices. \textbf{(3)} Quantization decreases energy efficiency for short and medium input–output pairs due to dequantization overheads dominating the computation. However, when input/output lengths grow, quantization improves energy efficiency.

\subsection{Load Intensity}\label{sec:char-load}

\begin{figure}[t]
  \centering
  \includegraphics[width=\linewidth]{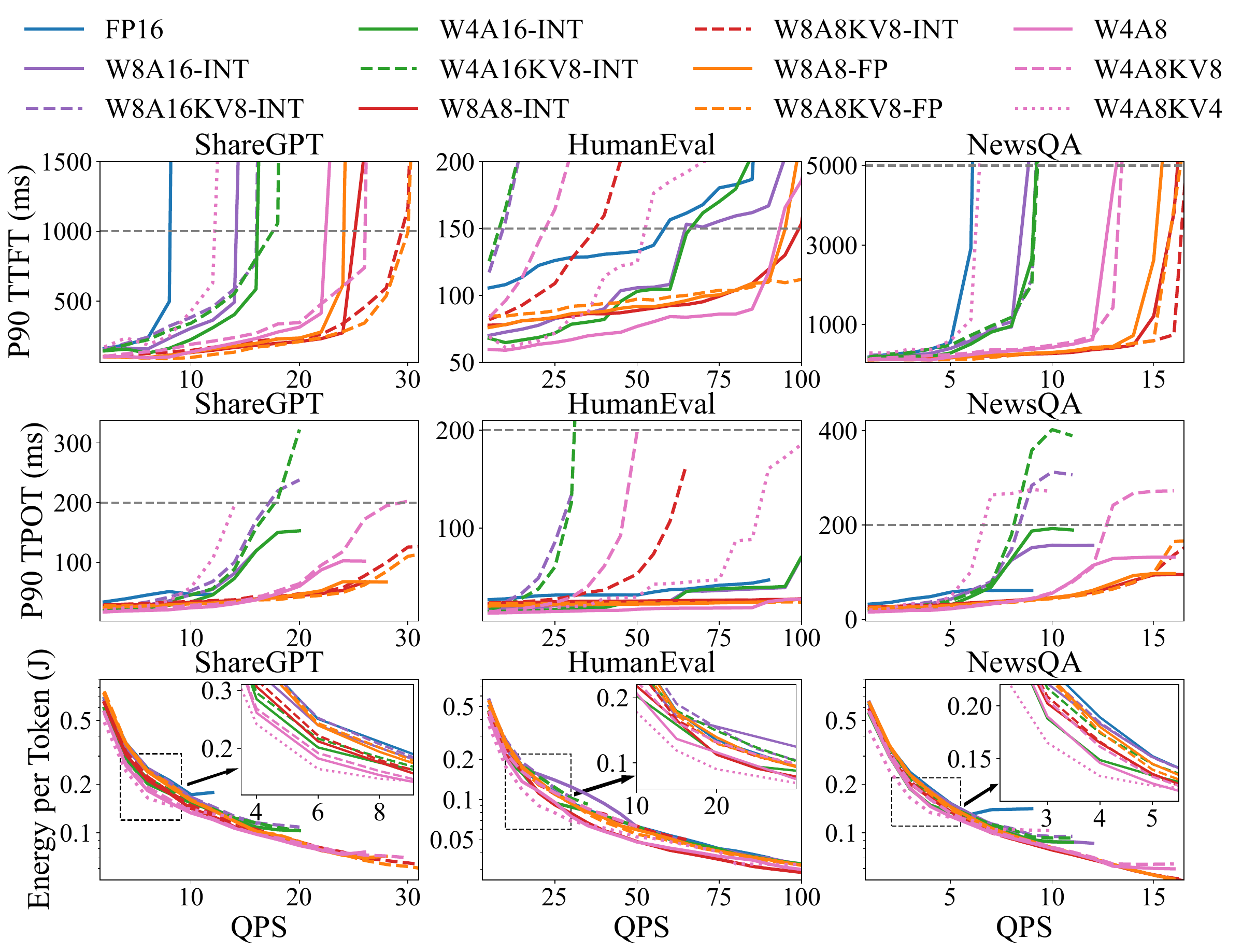}
  \caption{\textbf{Load influence (\Cref{sec:char-load}).} Latency and energy metrics of quantized 34B models under variable QPS. The horizontal black dash line represents latency SLO.}
  \label{fig:qps-curves}
\end{figure}

\begin{figure}[t]
  \centering
  \includegraphics[width=\linewidth]{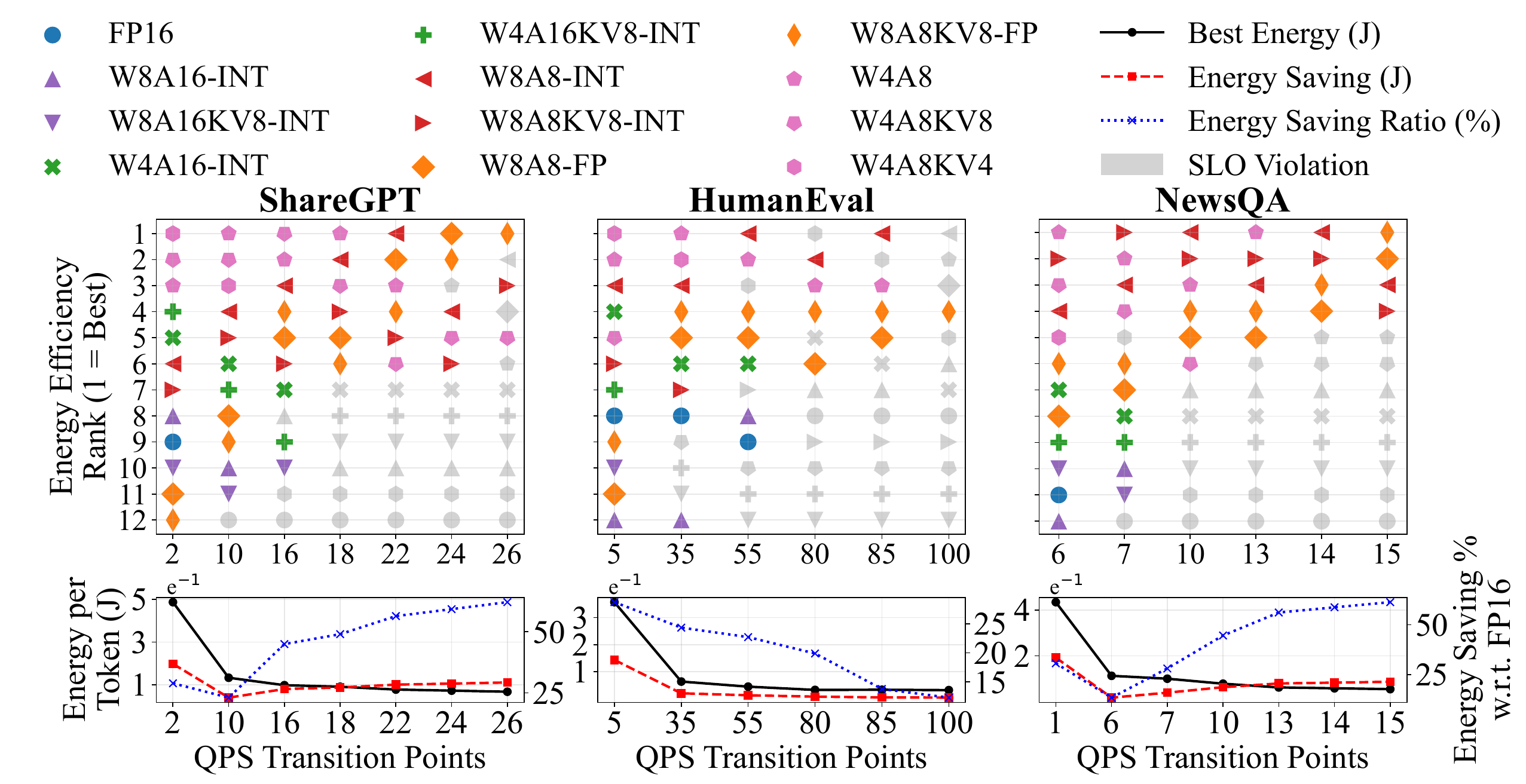}
  \caption{\textbf{Energy efficiency evolution  (\Cref{sec:char-load}).} Energy efficiency rank, best energy, best saving w.r.t. \FULLPREC{}, and best saving ratio of quantized 34B models under variable QPS. }
  \label{fig:qps-rank}
\end{figure}

\paragraph{Effect of quantization on saturation point.} \Cref{fig:qps-curves} shows P90 TTFT, P90 TPOT, and energy per token for different quantized 34B models across different request rates (QPS) on three tasks. The x-axes represent QPS (req/s), and the y-axes measure latency (ms) or energy per token (J). Across all three tasks, quantization generally pushes the saturation point higher than \FULLPREC{}, meaning that it can allow the system to sustain higher QPS before latency spikes. In both ShareGPT and NewsQA, nearly all quantized methods extend the saturation point, with more aggressive quantization yielding the largest improvements. However, HumanEval shows smaller gains; \emph{activation} methods still help, but KV cache compression can offset the benefits. 

\paragraph{Impact of traffic load on latency and energy.} Traffic load (request rate) strongly influences how quantization affects latency and energy efficiency as shown in~\Cref{fig:qps-curves}. \textbf{(1)} At low QPS, quantization provides modest latency benefits but already delivers noticeable energy savings, especially for lower-bit quantization. \textbf{(2)} As QPS increases, quantization yields greater latency reductions. Energy savings, shown in the bottom panel of \Cref{fig:qps-rank} and discussed in detail below, grow only when \FULLPREC{} saturates early; at low QPS, both the absolute and relative energy savings decrease. \textbf{(3)} KV cache compression can reduce or even negate latency gains under high loads, particularly in HumanEval, where short contexts make KV cache compression overhead more visible.

\paragraph{Shifts in energy-optimal and SLO-compliant configurations.} The combination of energy efficiency and SLO compliance changes with request rate. To illustrate this, we rank the energy-per-token values of all quantization methods and track how these rankings change with increasing traffic load, as shown in the top panel of \Cref{fig:qps-rank}. Methods that violate latency SLO constraints are shaded in gray. For clarity, we only display QPS \emph{transition points} where the optimal configuration changes, along with the corresponding best energy efficiency, absolute energy savings relative to \FULLPREC{}, and percentage savings at each transition shown in the bottom panel. We make the following observations. \textbf{(1)} The most energy-efficient SLO-compliant configuration depends on request rate: at low to mid QPS, heavily compressed models such as \AWQAC{} or \AWQACKV{} are often optimal, delivering 20–50\% energy savings while meeting latency SLOs. \textbf{(2)} Near saturation, lighter compression methods like \SMOOTHQUANT{} or \FPEightKV{} become preferable, as they maintain SLO compliance with slightly reduced energy savings. This shift underscores the importance of dynamically selecting quantization configurations based on load.

\noindent \cbox{\noindent \textbf{Finding \#2:} The benefits of quantization are highly sensitive to input/output length and load intensity. \textbf{(a)} Short outputs make dequantization and prefill overheads more noticeable, which can hurt TTFT, while long inputs make activation handling less efficient, leading to higher TPOT. \textbf{(b)} Load intensity shifts the balance of energy efficiency and latency, showing that no single configuration remains optimal across QPS. 

\noindent \textbf{Recommendation \#2:} \textbf{(a)} Workload-aware and load-adaptive scheduling are needed to dynamically select quantization methods based on prompt length, expected output length, and real-time system load. \textbf{(b)} Quantization methods should be developed to minimize prefill and dequantization overheads for short outputs and to better optimize activation handling for long inputs, ensuring more robust performance across diverse workloads.
}

\section{Parallelism}\label{sec:char-par}

\begin{figure}[t]
  \centering
  \includegraphics[width=\linewidth]{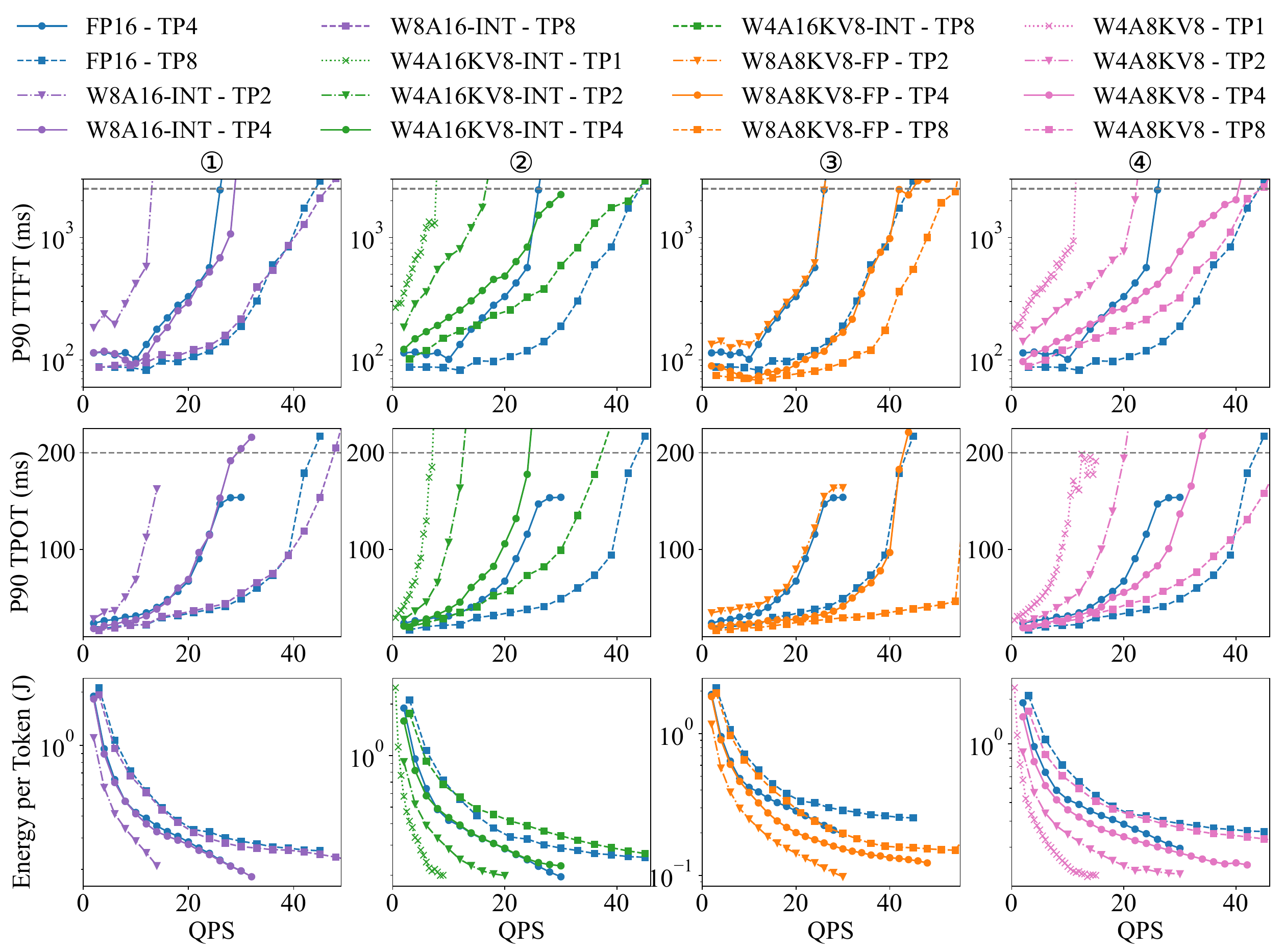}
  \caption{\textbf{Parallelism influence (\Cref{sec:char-par}).} Latency and energy trends of quantized Llama-2-70B models compared with FP16 across parallelism on the chatbot task. Trend group \ding{172}: weight-only; \ding{173}: weight-only with KV cache compression; \ding{174}: 8-bit weight and 8-bit activation with KV cache compression including \FPEight{}, \FPEightKV{}, \SMOOTHQUANT{}, W8A8KV8-INT, and \AWQAC{}; \ding{175}:4-bit weight and 8-bit activation with KV cache compression including \AWQACKV{} and \QSERVE{}.}
  \label{fig:parallelism}
\end{figure}

To serve LLMs of immense size, parallelism is needed to distribute computation across multiple GPUs. In this section, we examine how quantization interacts with parallelism. We focus on tensor parallelism (TP) in this study, running a 70B model with TP1/2/4/8 on H100 GPUs. TP is chosen because it performs better than pipeline parallelism in single-node settings, avoids pipeline bubbles and inter-stage synchronization, and is sufficient to host the model sizes we study on a 8 GPU node.
We also analyze data parallelism in~\Cref{sec:case2}. We identify four groups in the latency and energy trends of quantization methods under parallelism, and choose one representative per group to illustrate the key findings.

\Cref{fig:parallelism} reports P90 TTFT, P90 TPOT, and energy per token across selected quantization methods from each group. The x-axis represents QPS and the y-axis shows the metric values, with colors indicating quantization type and markers denoting the TP level. We make the following observations.
\textbf{(1)} For weight-only quantization (\ding{172}), adding more GPUs (TP2$\rightarrow$TP8) consistently reduces both TTFT and TPOT, but the relative latency and energy improvements from quantization remain fractional. This suggests that while parallelism alleviates compute bottlenecks, memory access and dequantization overheads still limit efficiency gains.
\textbf{(2)} For weight-only with KV cache compression (\ding{173}), scaling parallelism amplifies the overhead: latency often increases and energy efficiency degrades relative to \FULLPREC{}, especially at higher TP levels. This indicates that compression overhead interacts poorly with inter-GPU communication and sharded KV cache.
\textbf{(3)} For 8-bit weight and 8-bit activation with KV cache compression (\ding{174}), moderate parallelism amplifies the benefits. These methods consistently reduce both latency and energy across TP levels. Notably, \FPEightKV{} at TP4 achieves latency comparable to \FULLPREC{} at TP8 under SLO, demonstrating that quantization combined with moderate parallelism can replace heavier \FULLPREC{} scaling.
\textbf{(4)} For 4-bit weight and 8-bit activation with KV cache compression (\ding{175}), results are mixed. At low to mid QPS, latency often worsens compared to \FULLPREC{} due to additional compression/decompression overhead. However, near \FULLPREC{}'s saturation, these methods show relative latency gains and consistently better energy efficiency, suggesting that aggressive quantization becomes more competitive under heavy load.
\textbf{(5)} For 4-, 8-bit weight and 8-bit activation with KV cache compression (\ding{174},\ding{175}), high TP observes a diminishing return compared to moderate TP. Doubling GPUs from TP4 to TP8 only increases the saturation point by 20-35\% and leads to worse energy efficiency.

\noindent \cbox{\noindent \textbf{Finding \#3:} \textbf{(a)} Quantization interacts strongly with tensor parallelism: activation quantization scales well under moderate TP, but weight-only with KV compression incurs compounded latency and energy overheads. \textbf{(b)} Activation quantization (e.g., FP8-KV) at TP4 can match FP16 at TP8, achieving similar latency with fewer GPUs. \textbf{(c)} KV cache compression can undermine latency at higher TP due to additional communication and synchronization overheads.

\noindent \textbf{Recommendation \#3:} \textbf{(a)} The effectiveness of quantization depends on both precision and parallel execution. Parallelism-aware quantization is needed to align dequantization and KV handling with sharding layouts. \textbf{(b)} Quantization can act as a substitute for aggressive TP scaling. We can use quantization as a scaling lever in scheduling to reduce hardware demand and energy consumption.
}

\section{Hardware Level Analysis}\label{sec:char-hw}

\begin{figure}[t]
  \centering
  \includegraphics[width=\linewidth]{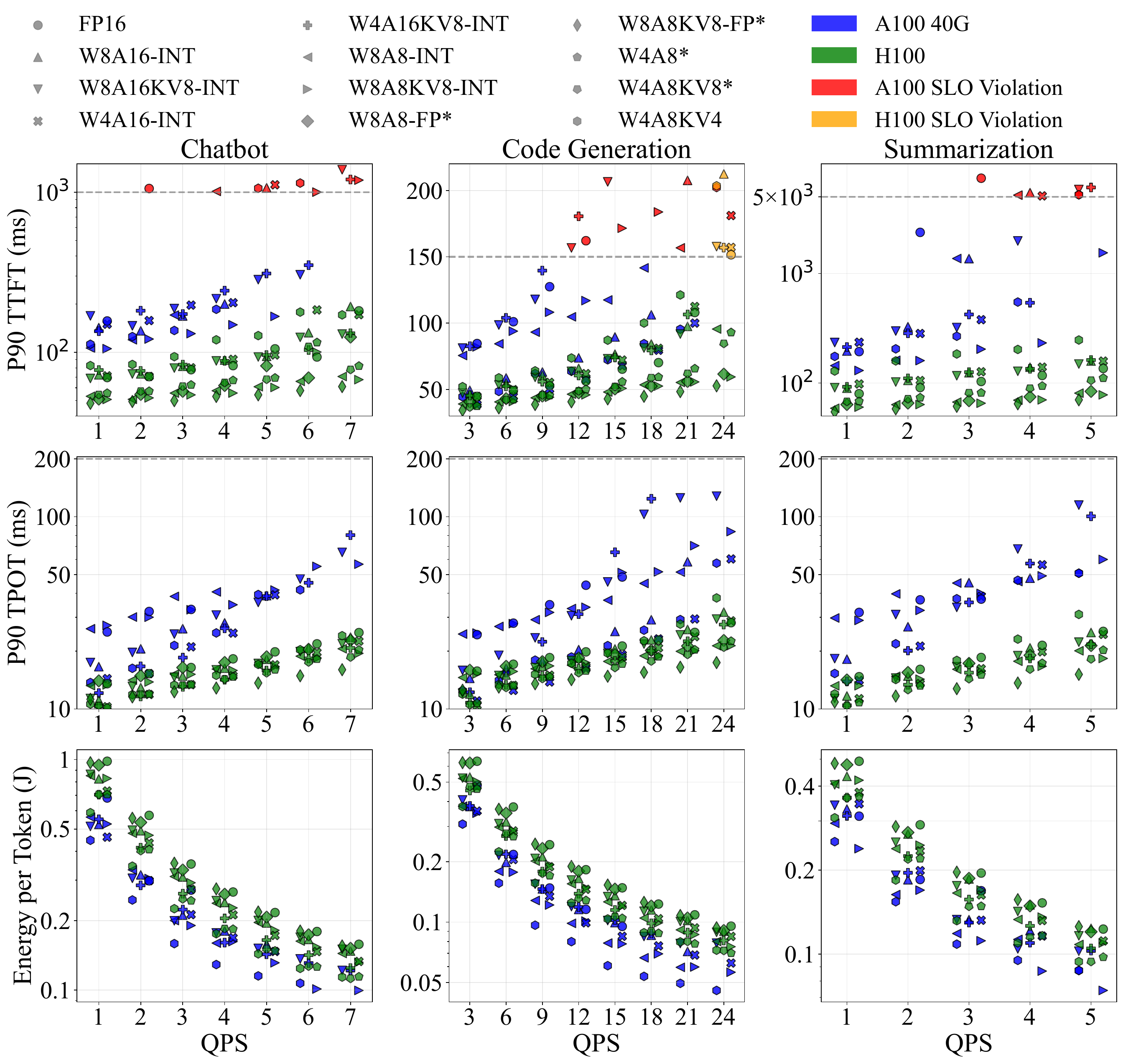}
  \caption{\textbf{Hardware influence (\Cref{sec:char-hw}).} Latency and energy metrics of quantized 13B models on H100 and A100 GPUs within A100's saturation range. Methods marked by * are only available on H100 for FP8 compute compatibility.}
  \label{fig:hw-pareto}
\end{figure}

\begin{table}[t]
\footnotesize
  \centering
  \caption{Hardware specifications of GPUs used in this study.}
  \label{tab:hw-specs}
  \begin{tabular}{llllll}
    \toprule
    \textbf{GPU}  & \textbf{Memory}   & \textbf{FP16} &\textbf{INT8} & \textbf{Memory} & \textbf{TDP} \\ 
     & \textbf{Capacity} & \textbf{TFLOPS} & \textbf{TOPS}  &  \textbf{Bandwidth} & \\ \midrule
    A100~\cite{nvidia_a100_datasheet} &40 GB         &  624  & 1248  & 1.6 TB/s & 400 W\\
    H100~\cite{nvidia_h100_datasheet}  & 80 GB       &  1979  & 3958 & 3.35 TB/s & 700 W\\
    \bottomrule
  \end{tabular}
\end{table}

Beyond application- and system-level analysis, we also examine how hardware platforms influence the behavior of different quantization methods. Specifically, we compare NVIDIA A100 and H100 GPUs, with their specifications in \Cref{tab:hw-specs}. To minimize communication and parallelism effects, we run the Llama-2 13B model on a single GPU in each case, as one device is sufficient to host the model.

\Cref{fig:hw-pareto} reports P90 TTFT, P90 TPOT, and energy per token across quantization methods on both GPUs. The x-axis represents QPS and the y-axis shows the metric values, with colors indicating GPU type and markers denoting quantization method. For visual clarity, the QPS values of different quantization methods are slightly offset, though they correspond to the same request rate. We make the following observations. \textbf{(1)} Overall, A100 exhibits higher latency than H100, even with quantization applied. Quantized models on A100 still saturate earlier than FP16 on H100, reflecting differences in raw compute capability. \textbf{(2)} Memory capacity also plays a role: the effective VRAM available for KV cache on A100 running a \AWQWO{} 13B model ($\approx$30 GB) is about 2.5$\times$ that of running a FP16 13B ($\approx$12 GB), but still significantly smaller than running a \FULLPREC{} 13B on H100 ($\approx$53 GB), limiting the maximum parallelism that can be achieved. \textbf{(3)} A100 shows better energy efficiency at low to mid QPS. Compared to FP16 on H100, the same quantization method delivers on average 9.6–35.6\% greater energy savings on A100. This difference stems from the higher TDP of H100, which boosts performance but also increases power consumption.

\noindent \cbox{\noindent \textbf{Finding \#4:} \textbf{(a)} Hardware architecture shapes the tradeoffs of quantization: newer GPUs like H100 improve latency and scalability, but older GPUs like A100 may deliver higher energy efficiency under moderate loads. \textbf{(b)} Quantization helps alleviate the memory capacity bottleneck, especially on older hardware. But beyond that, but system performance and saturation also depend on compute capability.

\noindent \textbf{Recommendation \#4:} Hardware-aware scheduling is needed to select quantization configurations based on both workload characteristics and GPU architecture.
}


\section{Optimizations for LLM Quantization Serving Systems}\label{sec:opt}

Building on the characterization results, this section presents three optimization case studies for quantization-enabled LLM serving clusters, motivated by real-world deployment needs and guided by model–system–hardware co-design.

\subsection{Saturation Point Prediction}\label{sec:case1}

To optimize the energy efficiency of a quantization-enabled LLM serving cluster, we want to identify the \emph{saturation point}, i.e., the maximum QPS each instance can sustain while meeting latency SLOs. This point determines usable throughput for scheduling and capacity planning. Exhaustive profiling provides accurate saturation points at high cost. We therefore ask: \emph{can machine learning models predict the saturation point for unseen configurations, reducing the need for profiling?}

\paragraph{Data sources and setup.} We leverage three sources of benchmarking data from prior profiling efforts:
\begin{enumerate}
    \item H100 benchmarking data covering 7B/13B/34B/70B models, ShareGPT/HumanEval/NewsQA workloads, tensor parallelism (TP) levels 1/2/4/8, and all 12 methods (277 data points).
    \item A100 benchmarking data with the same model sizes and datasets as H100, but restricted to TP levels 1/2/4 and 8 non-FP8 quantization methods (172 data points).
    \item Synthetic random-input workloads for the 34B TP1 model on H100, where we systematically control input and output length (192 data points).
\end{enumerate}
Each record is represented as a feature tuple: (\emph{model size, quantization method, GPU type, input length, output length}), and the prediction target is the measured saturation point. We train XGBoost~\cite{chen2016xgboost} regressors under different train-test split schemes to understand predictability.

\paragraph{Experiments and findings.} We experiments under the following three scenarios to understand predictability.
\begin{enumerate}
    \item \textbf{General learnability (random splits).} We begin with random data splits to assess the fundamental learnability of the prediction task. When using all available data, the model achieves a mean absolute percentage error (MAPE) of 31.1\% with an 8:2 train-test split, and 33.0\% with a 9:1 split. Excluding the HumanEval dataset—whose unusually long outputs produce saturation points exceeding 100 req/s and heavily skew the target distribution—reduces error to 18.9\%. A similar improvement is observed when restricting evaluation to H100 data alone (18.8\%), indicating greater consistency within a single hardware domain. Further narrowing to random-input workloads yields even better stability, with MAPE decreasing to 16.2\%. The lowest error, 14.9\%, is achieved when combining both restrictions: H100-only data while excluding HumanEval. Overall, these results suggest that heterogeneity across datasets and hardware introduces distribution shift, but within a constrained domain, saturation point prediction is feasible with moderate accuracy.
    \item \textbf{Unseen request lengths.} We evaluate extrapolation by excluding specific input or output lengths during training. Excluding input lengths yields poor generalization (MAPE 34.9–85.3\%), and excluding output lengths produces similar errors (16.8–69.5\%). Short outputs (64/128 tokens) are predicted more accurately due to overlap with natural workloads (ShareGPT/HumanEval). Incorporating higher TP (TP$>$1) slightly increases error by 1–10\%, indicating parallelism confounds length-dependent saturation behavior. The lowest error (14.9\% MAPE) occurs when restricting to H100 data and excluding HumanEval. Overall, interpolation across lengths is feasible, but extrapolation to unseen lengths remains unreliable. 
    \item \textbf{Cross-GPU transfer.} We examine whether knowledge can transfer across GPU types. Using H100 data with partial A100 data (HumanEval, NewsQA) to train and testing on A100 ShareGPT results in a high error (MAPE 73.1\%). Alternative configurations, such as swapping the train–test datasets (e.g., training on A100 ShareGPT and testing on HumanEval/NewsQA) or augmenting the training set with random-input workloads, do not improve accuracy and can yield errors exceeding 100\%. These results reveal a significant gap between H100 and A100 GPUs. Saturation behavior is hardware-specific; models trained on one GPU type is hard to generalize to another.
\end{enumerate}

\paragraph{Takeaway.} Learning-based prediction of saturation points is feasible within homogeneous data regimes (same GPU type, similar request lengths, consistent datasets), where errors can be reduced below 15\%. However, predictions fail across domains—especially between different GPUs or unseen request lengths—revealing strong domain gaps. This underscores that profiling remains indispensable for robust system design, particularly when optimizing energy efficiency across heterogeneous hardware or diverse workloads.

\subsection{Energy-Optimal Configuration with Data Parallelism}\label{sec:case2}

While we did not analyze data parallelism in~\Cref{sec:char-par}, we examine the role of data parallelism in improving energy efficiency here by evaluating system configurations across different model sizes and datasets. For each configuration defined by a pair of quantization method and tensor parallelism degree, we sweep the QPS range from 0 up to the saturation point of a single instance under tensor parallelism. When the target QPS exceeds this saturation point, we provision $\lceil \frac{x}{x_s}\rceil$ instances to meet the demand, distributing load evenly across instances. This load balancing is crucial because the energy–QPS curve is convex and monotonically decreasing (e.g., \Cref{fig:qps-curves}), allowing us to maximize energy efficiency. By repeating this process across all QPS values, we can identify the energy-optimal configuration for each workload.

\paragraph{Findings.} We summarize the findings as follows:
\begin{itemize}
    \item \textbf{Data and tensor parallelism tradeoffs.} For the profiled model sizes and datasets, data parallelism combined with lower tensor parallelism often outperforms a single high tensor parallelism H100 instance in terms of energy efficiency. This indicates that, on H100, tensor parallelism does not necessarily scale quantization benefits. 
    \item \textbf{Cross-GPU comparison (A100 vs. H100).} On A100, however, combining data and tensor parallelism can occasionally approach or even surpass the energy efficiency of an H100 configuration. For example, on the 34B HumanEval workload, an A100 TP2 configuration rivals the energy efficiency of H100 TP1 under moderate load. 
\end{itemize}

\paragraph{Takeaway.} Determining energy-optimal system configurations is inherently non-trivial. Simple rules—such as preferring tensor parallelism over data parallelism or always using newer GPUs—can lead to suboptimal outcomes.

\subsection{Energy–Quality Tradeoffs}

To better understand the energy–quality tradeoffs in deploying quantized models under high traffic, we synthesize a cluster-level request trace and compare three simple strategies for selecting system configurations (\emph{model size, quantization method, GPU type, tensor parallelism (TP), data parallelism (DP)}). The trace includes four request types: \texttt{chat-S}, \texttt{chat-R}, \texttt{code generation}, and \texttt{summarization}, each with specific latency and quality SLOs (TTFT/TPOT/Quality Score: 1s/0.2s/55, 3s/0.2s/50, 0.15s/0.2s/35, 5s/0.2s/16). Each request type is assumed to have a dedicated resource pool. We generate request lengths by sampling the Azure LLM trace~\cite{patel2024splitwise}, aligning them with benchmarking datasets and scaling the QPS to over 100 req/s 
to simulate cluster-level traffic, with a \texttt{chat-S}: \texttt{chat-R} ratio of 4:1. Based on profiling, we map request types to fixed model sizes: \texttt{chat-S} to 13B, \texttt{chat-R} to 70B, \texttt{code generation} to 34B, \texttt{summarization} to 13B.

We evaluate three strategies for system configurations:
\begin{enumerate}
    \item \textbf{\FULLPREC{}-Only:} Use FP16 models, calculate TP and DP levels to meet SLOs, pick the energy-optimal \emph{(GPU, DP, TP)}.
    \item \textbf{Quality-First:} Select the quantization method with highest output quality, then compute TP and DP to meet SLOs, finally pick energy-optimal \emph{(GPU, DP, TP)}.
    \item \textbf{Energy-First:} Select the \emph{(quantization method, GPU, TP)} with lowest energy per token at its saturation point, then compute the DP to meet SLOs. 
\end{enumerate}
For each timestamp in the synthetic trace, we apply the three strategies and record the total GPUs used, SLO attainment, and cluster-level energy per token.

\begin{table}[t]
\footnotesize
    \centering
    \caption{Energy-quality tradeoffs for three strategies.}
    \label{tab:res-case3}
    \begin{tabular}{llll}
    \toprule
        \textbf{Strategy} & \textbf{Avg \# GPUs} & \textbf{SLO attainment (\%)} & \textbf{Energy/Token}\\
        \midrule
       FP16-Only  & 45  & 100 & 0.128 J\\
       Quality-First & 53 & 100 & 0.137 J\\
       Energy-First & 24 & 38.6 & 0.062 J\\
       \bottomrule
    \end{tabular}

\end{table}

\paragraph{Findings.} \Cref{tab:res-case3} shows the results. Prioritizing quality increases GPU allocation and energy consumption due to dequantization overhead, while prioritizing energy can degrade output quality. For example, Energy-First chooses the 13B W8A8KV8-INT model for \texttt{chat-S} requests, violating the quality SLO: the average score drops from 60 (\FULLPREC{}) to 50.48, a 16\% decline. Since \texttt{chat-S} represents 50–60\% of the traffic, this illustrates the critical need to balance energy efficiency with quality preservation in real-world LLM serving.

\paragraph{Takeaway.} Energy–quality tradeoffs exist. Achieving the best energy-quality tradeoffs while meeting SLOs requires adaptive configuration strategies that dynamically balance model precision, parallelism, and resource allocation.

\section{Limitation Discussion}

Despite our best efforts to analyze state-of-the-art LLM quantization methods, our study cannot encompass all possible approaches. The main limitations are: 
\textbf{(1)} We focus on 8-bit and 4-bit quantization, and do not cover lower-bit formats such as 2-bit~\cite{chee2023quip}, as well as emerging mixed-precision~\cite{tao2025moqae,chen2024mixq} and adaptive schemes~\cite{ou2024adaptive}. These are not included because they are unsupported by our chosen inference engine for study, and their adoption in practice remains uncertain. 
\textbf{(2)} Our evaluation uses open-source Llama models with the TensorRT-LLM inference engine, which may limit applicability to other model families or serving frameworks. 
\textbf{(3)} Our hardware scope is restricted to NVIDIA H100 and A100 GPUs, without considering other NVIDIA GPU generations or alternative AI accelerators.
\textbf{(4)} We do not consider the impact of other inference optimizations such as chunked prefill~\cite{agrawal2024taming} and disaggregated serving~\cite{zhong2024distserve,patel2024splitwise,stojkovic2025dynamollm}, which are orthogonal to quantization but may interact in interesting ways (e.g., through communication overhead).
\textbf{(5)} Our benchmarking tasks do not consider very long-context workloads such as repository-level code completion ($\geq$8K tokens)~\cite{cheng2024dataflow} that are increasingly common. 
\textbf{(6)} For parallelism, we do not consider pipeline~\cite{ma2024hpipe}, 3D~\cite{chen2023ee}, or other parallelism techniques.
\textbf{(7)} On the model side, we do not consider the GGUF format~\cite{gguf_spec}, which supports lower-bit quantization and is popular for personal or edge devices, nor do we evaluate newer models that adopt native mixed-FP8 training rather than relying solely on post-training quantization.

\section{Conclusion}

This paper presents an online profiling tool and a joint performance, energy, and quality characterization of LLM quantization, laying the groundwork for model, system, hardware co-design for quantization-enabled LLM serving at scale.

\bibliographystyle{ACM-Reference-Format}
\bibliography{reference}

\end{document}